\documentstyle{article}    

\def\PPh#1{\setbox0\hbox{$#1\rm I$}\mathord{\vcenter{\ialign{$#1\rm##$\cr
I\cr\noalign{\nointerlineskip \vskip-0.541\ht0}P\cr}}}}

\def\Ph{{\mathpalette\PPh{}}}
\def\Ed{{\mathord{\mkern5mu\mathaccent"7020{\mkern-5mu\partial}}}}

\newcommand{\Edt}{\tilde{\Ed}}
\newcommand{\Pht}{\tilde{\Ph}}

\let\Eqnarray=\eqnarray
\renewcommand{\eqnarray}{\arraycolsep=0.1675em \Eqnarray}

\newcommand{\art}[6]{{\rm #1, \rm #2, \it #3 \bf #4 \rm (#5), \mbox{#6}.}}
\newcommand{\artnotitle}[5]{{\rm #1, \it #2 \bf #3 \rm (#4), \mbox{#5}.}}
\newcommand{\artnopage}[5]{{\rm #1, \rm #2, \it #3 \bf #4 \rm (#5).}}

\newcommand{\book}[3]{{\rm #1, \it #2, \rm #3.}}

\newtheorem{proposition}{Proposition}[section]

\newtheorem{definition}[proposition]{Definition}
\newtheorem{example}[proposition]{Example}

\let\Definition=\definition
\renewcommand{\definition}{\Definition \rm}
\let\Example=\example
\renewcommand{\example}{\Example \rm}

\newcommand{\ch}{{\raise 0.69mm \hbox{$\chi$}}}

\begin{document}

\centerline{\Large{\bf  INTEGRATION OF THE GHP EQUATIONS
}}

\centerline{\Large{\bf  IN SPACETIMES ADMITTING
}}

\centerline{\Large{\bf  A GEODESIC SHEAR-FREE
}}

\centerline{\Large{\bf  EXPANDING NULL CONGRUENCE
}}

\

\centerline{{\bf Fredrik Andersson}}

\centerline{ \it Department of Mathematics,
 Link\"oping University,}

\centerline{\it S581 83 Link\"oping,}

\centerline{\it Sweden. }

\centerline{Email: frand@mai.liu.se}

\

\centerline{{\bf Abstract}}

\

We perform a complete $\rho$-integration of the GHP equations for all 
spacetimes that admit a geodesic shear-free expanding null congruence,
whose Ricci spinor is aligned to the congruence and whose Ricci scalar
is constant. We also deduce the system of GHP equations after the
integration is completed, and discuss a few applications.

\


\newpage

\section{Introduction}

\subsection{Conventions}

We will use spacetime definitions and conventions from \cite{PR1}. In 
particular this means that the metric $g_{ab}$ is assumed to have 
signature $(+--\,\,-)$. We will use spinors for our calculations, but
as all results are local in nature there is no need to postulate the 
existence of a global spinor structure on spacetime. All spinor 
dyads $(o^{A},\iota^{A})$ will be assumed to be normalized i.e., 
$o_{A}\iota^{A}=1$. $\nabla_{AA'}$ denotes the Levi-Civita connection i.e.,
the uniquely defined metric and torsion-free (symmetric) connection on
spacetime.

\subsection{Outline}

When looking for exact solutions of Einstein's field equations in 
NP- or GHP-formalism, a common approach is to assume the existence
of a geodesic shear-free expanding null congruence. In vacuum spacetimes,
the Goldberg-Sachs theorem tells us that this condition is necessary and
sufficient for the Weyl curvature spinor to be algebraically special. In
this paper we will relax the vacuum condition and instead assume that
the GHP-components of the Ricci spinor satisfy
$$
 \Phi_{00}=\Phi_{01}=\Phi_{02}=0
$$
or equivalently
$$
 \Phi_{ABA'B'}o^{A'}o^{B'}=0
$$
where $l^{a}=o^{A}o^{A'}$ denotes the geodesic shear-free expanding
congruence. This assumption is still sufficient to ensure algebraic 
speciality of the Weyl spinor. For technical reasons we will also 
assume
$$
 \Lambda={\rm constant}.
$$
We then show that all GHP equations containing $\Ph$ can be integrated
explicitly, something which is done in \cite{Held1} for the vacuum
case.\footnote{Kerr \cite{Kerr} has also integrated the GHP equations 
using this method for the case $\Phi_{ABA'B'}=0$ i.e., Einstein 
spacetimes.} We also show that the remaining GHP equations take on a very
simple appearance after the integration procedure is completed and discuss
redundancy in the final system of equations.

It is worth noting that a similar integration in NP-formalism for a 
non-vacuum subclass of these spacetimes is performed in \cite{Ludwig1} 
and \cite{Ludwig2} and, under even more restrictive assumptions, this 
NP integration have been performed in e.g., \cite{Talbot}, 
\cite{Lind}, \cite{TW1} and \cite{TW2}
 
\section{Spacetimes admitting a geodesic shear-free expanding null
congruence}

\subsection{The general case}

We will consider spacetimes admitting a geodesic, shear-free null
congruence $l^{a}=o^{A}o^{A'}$ and whose Ricci spinor satisfies the 
condition
\begin{equation}
 \Phi_{ABA'B'}o^{A}o^{B}=0.
\end{equation}
Take $o^{A}$ as the first spinor of a spinor dyad. In GHP-formalism 
the above conditions are equivalent to
\begin{equation}
 \Phi_{00}=\Phi_{01}=\Phi_{02}=0\quad,\quad\kappa=\sigma=0
\end{equation}
By the Goldberg-Sachs theorem we obtain
\begin{equation}
 \Psi_{0}=\Psi_{1}=0
\end{equation}
so the spacetime is algebraically special.

In addition we will assume that $\rho\neq 0$ so that $l^{a}$ is 
expanding. Then we can use a null rotation about $o^{A}$ to achieve
$\tau=0$, and the Ricci equations \cite{PR1} then imply that also
\begin{equation}
 \tau'=\sigma'=0.
\end{equation}
We introduce Held's \cite{Held1} modified operators which can be written
\begin{equation}
 \Edt=\frac{1}{\bar{\rho}}\Ed \quad ,\quad\Edt'=\frac{1}{\rho}\Ed'
 \quad ,\quad \Pht'=\Ph'+\frac{p}{2\rho}(\Psi_{2}+2\Lambda)+
 \frac{q}{2\bar{\rho}}(\bar{\Psi}_{2}+2\Lambda)
\end{equation}
in this dyad. Note that the definition of $\Pht'$ is slightly modified 
from Held's in our non-vacuum case. The purpose of using Held's modified 
operators is simply to reduce the length of calculations; in 
particular the new operators have the nice properties
\begin{equation}
 \Bigl[\Ph,\Edt\Bigr]=\Bigl[\Ph,\Edt'\Bigr]=0
\end{equation}
and
\begin{equation}
	\Bigl[\Ph,\Pht'\Bigr]\eta=\bigl[-\frac{1}{2\rho}(\Psi_{2}+2\Lambda)-
	\frac{1}{2\bar{\rho}}(\bar{\Psi}_{2}+2\Lambda)\bigr]\Ph\eta
	\label{}
\end{equation}
so if $\eta^{\circ}$ satisfies $\Ph\eta^{\circ}=0$ (a degree sign will,
throughout the paper, be used to denote a quantity that is killed by $\Ph$)
then
$$
\Ph \Edt'\eta^{\circ}=\Bigl[\Ph,\Edt'\Bigr] \eta^{\circ}=0
$$
and the same result is true if $\Edt'$ is replaced with $\Edt$ or $\Pht'$.

From the Ricci equations \cite{PR1} we obtain the equations
\begin{equation}
 \Ph \rho=\rho^2 \,,\; \Edt \rho=0
\end{equation}
The first of these equations can be used to `$\rho$-integrate' some of
the Ricci- and Bianchi equations. However, we first need to calculate
$\Edt'\rho$. Therefore, define the twist of the congruence $l^{a}$ as
\begin{equation}
 \Omega^{\circ}=\frac{1}{\bar{\rho}}-\frac{1}{\rho}
\end{equation}
We then obtain
$$
 \Ph \Omega^{\circ}=-\frac{\bar{\rho}^2}{\bar{\rho}^2}+
 \frac{\rho^2}{\rho^2}=0
$$
so the notation is consistent. We also obtain
\begin{equation}
 \Edt' \rho=\rho^2 \Edt' \Omega^{\circ}.
\end{equation}
The Ricci- and Bianchi equations \cite{PR1} involving $\Ph$, but not 
$\Phi_{22}$ are
\begin{eqnarray}
 \Ph\rho' & = & \bar{\rho}\rho'-\Psi_{2}-2\Lambda \nonumber \\
 \Ph\kappa' & = & -\Psi_{3}-\Phi_{21} \nonumber \\
 \Ph (\Psi_{2}+2\Lambda) & = & 3\rho \Psi_{2}+2\rho \Phi_{11} 
 \nonumber \\Ê\Ph \Psi_{3}-\Ph \Phi_{21}-\rho \Edt'(\Psi_{2}+2\Lambda)
 & = & 2\rho \Psi_{3}-2\bar{\rho}\Phi_{21} \nonumber \\Ê\Ph \Psi_{4}
 -\rho \Edt' \Psi_{3}-\rho \Edt' \Phi_{21} & = & \rho \Psi_{4} 
 \nonumber \\Ê\Ph (\Phi_{11}+3\Lambda) & = & 2(\rho+\bar{\rho})
 \Phi_{11} \nonumber \\ \Ph \Phi_{21}-\rho \Edt'(\Phi_{11}-3\Lambda)
 & = & (\rho+2\bar{\rho})\Phi_{21}
\end{eqnarray}
In order to integrate these with respect to $\Ph$ we need to know the 
dependence of the Ricci scalar $\Lambda$ on $\Ph$. Therefore we will 
make the additional assumption that $\Lambda$ is constant\footnote{It 
would in fact be sufficient to assume that $\Lambda$ is killed by 
$\Ph$, $\Ed$ and $\Ed'$, but unless $\rho=\bar{\rho}$ the commutators 
would imply that also $\Ph'\Lambda=0$, so we choose to ignore this 
possibility. It may also be sufficient to assume some other explicit 
$\Ph-$, $\Ed-$ and $\Ed'-$dependence of $\Lambda$, $\Psi_{2}$, or 
$\Phi_{11}$.}. Then these equations reduce to
\begin{eqnarray}
 \Ph\rho' & = & \bar{\rho}\rho'-\Psi_{2}-2\Lambda \nonumber \\
 \Ph\kappa' & = & -\Psi_{3}-\Phi_{21} \nonumber \\
 \Ph \Psi_{2} & = & 3\rho \Psi_{2}+2\rho \Phi_{11} \nonumber \\
 \Ph \Psi_{3}-\Ph \Phi_{21}-\rho \Edt' \Psi_{2} & = & 2\rho \Psi_{3}
 -2\bar{\rho}\Phi_{21} \nonumber \\Ê\Ph \Psi_{4}-\rho \Edt' \Psi_{3}
 -\rho \Edt' \Phi_{21} & = & \rho \Psi_{4} \nonumber \\Ê\Ph \Phi_{11}
 & = & 2(\rho+\bar{\rho})\Phi_{11} \nonumber \\ \Ph \Phi_{21}-\rho
 \Edt' \Phi_{11} & = & (\rho+2\bar{\rho})\Phi_{21}
\end{eqnarray}
The sixth of these equations is equivalent to
$$
 0=\frac{1}{\rho^2\bar{\rho}^2}\Ph \Phi_{11}-\Bigl(\frac{2}
 {\rho \bar{\rho}^2}+\frac{2}{\rho^2 \bar{\rho}}\Bigr)\Phi_{11}=
 \Ph \Bigl(\frac{\Phi_{11}}{\rho^2 \bar{\rho}^2}\Bigr)
$$
Thus, we obtain
\begin{equation}
	\Phi_{11}=\rho^2\bar{\rho}^2\Phi_{11}^{\circ}
	\label{}
\end{equation}
where, as usual $\Ph \Phi_{11}^{\circ}=0$. Then the third equation can 
be written
$$
 0=\frac{1}{\rho^3}\Ph \Psi_{2}-\frac{3}{\rho^2}\Psi_{2}-2\bar{\rho}^2
 \Phi_{11}^{\circ}=\Ph \Bigl(\frac{\Psi_{2}}{\rho^3}-2\bar{\rho}
 \Phi_{11}^{\circ}\Bigr).
$$
This gives us
\begin{equation}
	\Psi_{2}=\rho^3\Psi_{2}^{\circ}+2\rho^3\bar{\rho}\Phi_{11}^{\circ}.
	\label{}
\end{equation}
In a similar way, the remaining of the above equations can be 
integrated, to give
\begin{eqnarray}
 \rho' & = & \bar{\rho}\rho'^{\circ}-\frac{1}{2}(\rho^2+\rho\bar{\rho})
 \Psi_{2}^{\circ}-\rho^2\bar{\rho}\Phi_{11}^{\circ}+\frac{1}{\bar{\rho}}
 \Lambda \nonumber \\ \kappa' & = & \kappa'^{\circ}-\rho\Psi_{3}^{\circ}
 -\frac{1}{2}\rho^2\Edt'\Psi_{2}^{\circ}-\frac{1}{2}\rho^3\Psi_{2}^{\circ}
 \Edt'\Omega^{\circ}-\rho\bar{\rho}\Phi_{21}^{\circ}-\rho^2\bar{\rho}
 \Edt'\Phi_{11}^{\circ}-\rho^3\bar{\rho}\Phi_{11}^{\circ}\Edt'
 \Omega^{\circ} \nonumber \\
 \Phi_{21} & = & \rho\bar{\rho}^2\Phi_{21}^{\circ}+\rho^2\bar{\rho}^2
 \Edt' \Phi_{11}^{\circ}+\rho^3\bar{\rho}^2\Phi_{11}^{\circ}\Edt'
 \Omega^{\circ} \nonumber \\ \Psi_{3} & = & \rho^2 \Psi_{3}^{\circ}
 +\rho^3\Edt' \Psi_{2}^{\circ}+\frac{3}{2}\rho^4\Psi_{2}^{\circ}\Edt'
 \Omega^{\circ}+\rho^2\bar{\rho}\Phi_{21}^{\circ}+2\rho^3\bar{\rho}\Edt'
 \Phi_{11}^{\circ}+3\rho^4\bar{\rho}\Phi_{11}^{\circ}\Edt' \Omega^{\circ}
 \nonumber \\ \Psi_{4} & = & \rho\Psi_{4}^{\circ}+\rho^2\Edt'
 \Psi_{3}^{\circ}+\frac{1}{2}\rho^3\bigl(\Edt'^2\Psi_{2}^{\circ}+2
 \Psi_{3}^{\circ}\Edt'\Omega^{\circ}\bigr)+\frac{1}{2}\rho^4\bigl(
 \Psi_{2}^{\circ}\Edt'^2\Omega^{\circ}+3\Edt'\Omega^{\circ}\Edt'
 \Psi_{2}^{\circ}\bigr) \nonumber \\ & & +\frac{3}{2}\rho^5\Psi_{2}^{\circ}
 (\Edt'\Omega^{\circ})^2+\rho^2\bar{\rho}\Edt'\Phi_{21}^{\circ}+\rho^3
 \bar{\rho}\bigl(\Edt'^2\Phi_{11}^{\circ}+\Phi_{21}^{\circ}\Edt'
 \Omega^{\circ}\bigr) \nonumber \\ & & +\rho^4\bar{\rho}\bigl(
 \Phi_{11}^{\circ}\Edt'^2\Omega^{\circ}+3\Edt'\Omega^{\circ}\Edt'
 \Phi_{11}^{\circ}\bigr)+3\rho^5\bar{\rho}\Phi_{11}^{\circ}(\Edt'
 \Omega^{\circ})^2. \label{SpinCoeffs}
\end{eqnarray}
The remaining Ricci- and Bianchi equations are
\begin{eqnarray}
 \Ph'\rho & = & \rho\bar{\rho}'-\Psi_{2}-2\Lambda \nonumber \\
 \Ph'\rho'-\Ed\kappa' & = & \rho'^2+\Phi_{22} \nonumber \\
 \Ed'\kappa' & = & -\Psi_{4} \nonumber \\
 \Ed'\rho' & = & (\bar{\rho}-\rho)\kappa'-\Psi_{3}+\Phi_{21} 
 \nonumber \\ \Ed\Psi_{2} & = & 2\rho\Phi_{12} \nonumber \\
 \Ph'\Psi_{2}-\Ed\Psi_{3}-\Ed\Phi_{21}+\Ph\Phi_{22} & = & 3\rho'
 \Psi_{2}+\bar{\rho}\Phi_{22}+2\rho'\Phi_{11} \nonumber \\
 \Ph'\Psi_{3}-\Ed\Psi_{4}-\Ph'\Phi_{21}+\Ed'\Phi_{22} & = & 4\rho'
 \Psi_{3}-3\kappa'\Psi_{2}-2\bar{\rho}'\Phi_{21}+2\kappa'\Phi_{11}
 \nonumber \\ \Ph\Phi_{22}+\Ph'\Phi_{11}-\Ed'\Phi_{12}-\Ed\Phi_{21}
 & = & (\rho+\bar{\rho})\Phi_{22}+2(\rho'+\bar{\rho}')\Phi_{11}
 \label{restBianRic}
\end{eqnarray} 
We will now use the expressions (\ref{SpinCoeffs}), to rewrite these
equations. Starting with the first one, we obtain
\begin{equation}
	\Pht'\rho=\rho^2\bar{\rho}'^{\circ}-\frac{1}{2}\rho^2\bar{\rho}
	\overline{\Psi}_{2}^{\circ}-\frac{1}{2}\rho^3\Psi_{2}^{\circ}-
	\rho^3\bar{\rho}\Phi_{11}^{\circ}+\frac{\rho}{\bar{\rho}}\Lambda.
	\label{}
\end{equation}
From this equation we obtain the relation
$$
 \Pht'\Omega^{\circ}=\bar{\rho}'^{\circ}-\rho'^{\circ},
$$
and by applying the $\Bigl[\Ed,\Ed'\Bigr]$-commutator to $\rho$ we 
obtain the useful relation
$$
 \Edt\Edt'\Omega^{\circ}=2\Omega^{\circ}\bar{\rho}'^{\circ}+\Psi_{2}^{\circ}
 -\overline{\Psi}_{2}^{\circ}.
$$
Similarly, the substitution of known expressions into the third, 
fourth and fifth equation of (\ref{restBianRic}) easily give us
\begin{eqnarray}
 \Edt'\kappa'^{\circ} & = & -\Psi_{4}^{\circ} \nonumber \\ \Edt'
 \rho'^{\circ} & = & -\Omega^{\circ}\kappa'^{\circ}-\Psi_{3}^{\circ}
 \nonumber \\ \Edt\Psi_{2}^{\circ} & = & 2\overline{\Phi}_{21}^{\circ}
\end{eqnarray}
As we have already calculated the derivative operators action on 
$\Omega^{\circ}$ we can replace their action on $\rho$ with their 
action on the real $(-1,-1)$-quantity $\frac{1}{\rho}+\frac{1}
{\bar{\rho}}$
\begin{eqnarray}
 \Ph \Bigl(\frac{1}{\rho}+\frac{1}{\bar{\rho}}\Bigr) & = & -2 \nonumber \\
 \Edt \Bigl(\frac{1}{\rho}+\frac{1}{\bar{\rho}}\Bigr) & = & \Edt\Omega^{\circ}
 \nonumber \\ \Edt'\Bigl(\frac{1}{\rho}+\frac{1}{\bar{\rho}}\Bigr) & = &
 -\Edt'\Omega^{\circ} \nonumber \\ \Pht'\Bigl(\frac{1}{\rho}+\frac{1}
 {\bar{\rho}}\Bigr) & = & -(\rho'^{\circ}+\bar{\rho}'^{\circ})+\rho
 \Psi_{2}^{\circ}+\bar{\rho}\overline{\Psi}_{2}^{\circ}+2\rho\bar{\rho}
 \Phi_{11}^{\circ}-\frac{2\Lambda}{\rho\bar{\rho}}. 
\end{eqnarray}
We will next use the last equation of (\ref{restBianRic}) to 
$\rho$-integrate for $\Phi_{22}$, but before that we give two of the
commutators that will be used in the integration
\begin{eqnarray}
 \Bigl[\Pht',\Edt'\Bigr] & = & \Bigl(-\frac{\kappa'^{\circ}}{\rho}+
 \Psi_{3}^{\circ}+\frac{1}{2}\rho\Edt'\Psi_{2}^{\circ}+\frac{1}{2}\rho^2
 \Psi_{2}^{\circ}\Edt'\Omega^{\circ}+\bar{\rho}\Phi_{21}^{\circ}+\rho
 \bar{\rho}\Edt'\Phi_{11}^{\circ} \nonumber \\ & & +\rho^2\bar{\rho}
 \Phi_{11}^{\circ}\Edt'\Omega^{\circ}\Bigr)\Ph+p\bigl(\kappa'^{\circ}+
 \Lambda\Edt'\Omega^{\circ}\bigr) \nonumber \\ \Bigl[\Edt,\Edt'\Bigr] & = &
 \Bigl(\frac{\bar{\rho}'^{\circ}}{\bar{\rho}}-\frac{\rho'^{\circ}}{\rho}+
 \frac{\rho}{2}\bigl(\frac{1}{\rho}+\frac{1}{\bar{\rho}}\bigr)\Psi_{2}^{\circ}
 -\frac{\bar{\rho}}{2}\bigl(\frac{1}{\rho}+\frac{1}{\bar{\rho}}\bigr)
 \overline{\Psi}_{2}^{\circ}+\Omega^{\circ}(\rho\bar{\rho}\Phi_{11}^{\circ}
 -\Lambda)\Bigr)\Ph \nonumber \\Ê& & +\Omega^{\circ}\Pht'+p\bigl(
 \rho'^{\circ}+\Omega^{\circ 2}\Lambda\bigr)-q(\bar{\rho}'^{\circ}
 +\Omega^{\circ 2}\Lambda\bigr). \label{twoComms}
\end{eqnarray}
The last equation of (\ref{restBianRic}) requires considerably more 
work to $\rho$-integrate than the ones integrated previously. However, 
after a very long calculation involving these commutators, we obtain
\begin{eqnarray}
 \Phi_{22} & = & \rho\bar{\rho}\Phi_{22}^{\circ}+\rho^2\bar{\rho}\bigl(\Edt'
 \overline{\Phi}_{21}^{\circ}-\frac{1}{2}\Pht'\Phi_{11}^{\circ}\bigr)
 +\rho\bar{\rho}^2\bigl(\Edt\Phi_{21}^{\circ}-\frac{1}{2}\Pht'\Phi_{11}^{\circ}
 \bigr) \nonumber \\Ê& & +\rho^3\bar{\rho}\overline{\Phi}_{21}^{\circ}\Edt'
 \Omega^{\circ}+\frac{1}{2}\rho^2 \bar{\rho}^2\bigl(\Edt\Edt'\Phi_{11}^{\circ}
 +\Edt'\Edt\Phi_{11}^{\circ}\bigr)-\rho\bar{\rho}^3\Phi_{21}^{\circ}\Edt
 \Omega^{\circ} \nonumber \\ & & +\rho^3\bar{\rho}^2\Edt'\Omega^{\circ}\Edt
 \Phi_{11}^{\circ}-\rho^2\bar{\rho}^3\Edt\Omega^{\circ}\Edt'\Phi_{11}^{\circ}
 -\rho^3\bar{\rho}^3\Phi_{11}^{\circ}\Edt\Omega^{\circ}\Edt'\Omega^{\circ}
 \label{twistPhi22}
\end{eqnarray}
Next we will look at the sixth equation of (\ref{restBianRic}). We 
start by subtracting it from the last equation of (\ref{restBianRic}) 
to get rid of the $\Ph\Phi_{22}$-term
\begin{equation}
	\Ph'\Phi_{11}-\Ed'\Phi_{12}-\Ph'\Psi_{2}+\Ed\Psi_{3}=-3\rho'\Psi_{2}
	+2\bar{\rho}'\Phi_{11}+\rho\Phi_{22}.
	\label{restBianRic6}
\end{equation}
By using the previous equations along with the
$\Bigl[\Edt,\Edt'\Bigr]$-commutator it becomes
\begin{equation}
 \Edt\Psi_{3}^{\circ}-\Pht'\Psi_{2}^{\circ}=\Phi_{22}^{\circ}
 \label{Bian1}
\end{equation}
It remains to check the second and seventh equation of 
(\ref{restBianRic}). Using the equations obtained so far, the second 
gives us
\begin{equation}
	\Pht'\rho'^{\circ}-\Edt\kappa'^{\circ}=\Lambda(2\Omega^{\circ}
	\rho'^{\circ}+\Psi_{2}^{\circ}-\overline{\Psi}_{2}^{\circ}).
	\label{}
\end{equation}
Checking the seventh equation of (\ref{restBianRic}) is a very long 
and tedious calculation where both of the commutators (\ref{twoComms})
are used. The end result is that
\begin{equation}
	\Edt\Psi_{4}^{\circ}-\Pht'\Psi_{3}^{\circ}=\Lambda\bigl(\Edt'
	\Psi_{2}^{\circ}-2\Phi_{21}^{\circ}-2\Omega^{\circ}\Psi_{3}^{\circ}
	\bigr)
	\label{Bian2}
\end{equation}
As could be expected, the system of equations that we have obtained 
contains considerable redundancy. As an example of this we note
that the imaginary part of equation (\ref{Bian1}) and the whole of 
equation (\ref{Bian2}) are actually consequences of the other 
equations.

\section{Summary and conclusions}

\subsection{Summary}

In this section we will collect the resulting equations in one place.

First of all, the equations for the GHP-operators acting on $\rho$
\begin{eqnarray}
 \Ph \rho & = & \rho^2 \nonumber \\ \Edt \rho & = & 0 \nonumber \\ \Edt'
 \rho & = & \rho^2 \Edt'\Omega^{\circ} \nonumber \\ \Pht'\rho & = & \rho^2
 \bar{\rho}'^{\circ}-\frac{1}{2}\rho^2\bar{\rho}\overline{\Psi}_{2}^{\circ}
 -\frac{1}{2}\rho^3\Psi_{2}^{\circ}-\rho^3\bar{\rho}\Phi_{11}^{\circ}+
 \frac{\rho}{\bar{\rho}}\Lambda
\end{eqnarray}
can be split into the relations
\begin{eqnarray}
 \Ph \Bigl(\frac{1}{\rho}+\frac{1}{\bar{\rho}}\Bigr) & = & -2 \nonumber \\
 \Edt \Bigl(\frac{1}{\rho}+\frac{1}{\bar{\rho}}\Bigr) & = & \Edt\Omega^{\circ}
 \nonumber \\ \Edt'\Bigl(\frac{1}{\rho}+\frac{1}{\bar{\rho}}\Bigr) & = &
 -\Edt'\Omega^{\circ} \nonumber \\ \Pht'\Bigl(\frac{1}{\rho}+\frac{1}
 {\bar{\rho}}\Bigr) & = & -(\rho'^{\circ}+\bar{\rho}'^{\circ})+\rho
 \Psi_{2}^{\circ}+\bar{\rho}\overline{\Psi}_{2}^{\circ}+2\rho\bar{\rho}
 \Phi_{11}^{\circ}-\frac{2\Lambda}{\rho\bar{\rho}} \nonumber \\ \Ph
 \Omega^{\circ} & = & 0 \nonumber \\ \Pht'\Omega^{\circ} & = &
 \bar{\rho}'^{\circ}-\rho'^{\circ} \label{rOmegaeqns}
\end{eqnarray}
From the commutators on $\rho$ we have the useful relation
\begin{equation}
 \Edt\Edt'\Omega^{\circ}=2\Omega^{\circ}\bar{\rho}'^{\circ}+\Psi_{2}^{\circ}
 -\overline{\Psi}_{2}^{\circ}.
	\label{}
\end{equation}
The curvature scalars and the spin coefficients are
\begin{eqnarray}
 \rho' & = & \bar{\rho}\rho'^{\circ}-\frac{1}{2}(\rho^2+\rho\bar{\rho})
 \Psi_{2}^{\circ}-\rho^2\bar{\rho}\Phi_{11}^{\circ}+\frac{1}{\bar{\rho}}
 \Lambda \nonumber \\ \kappa' & = & \kappa'^{\circ}-\rho\Psi_{3}^{\circ}
 -\frac{1}{2}\rho^2\Edt'\Psi_{2}^{\circ}-\frac{1}{2}\rho^3\Psi_{2}^{\circ}
 \Edt'\Omega^{\circ}-\rho\bar{\rho}\Phi_{21}^{\circ}-\rho^2\bar{\rho}
 \Edt'\Phi_{11}^{\circ}-\rho^3\bar{\rho}\Phi_{11}^{\circ}\Edt'
 \Omega^{\circ} \nonumber \\ \Psi_{2} & = & \rho^3\Psi_{2}^{\circ}+2\rho^3
 \bar{\rho}\Phi_{11}^{\circ} \nonumber \\ \Psi_{3} & = & \rho^2
 \Psi_{3}^{\circ}+\rho^3\Edt' \Psi_{2}^{\circ}+\frac{3}{2}\rho^4
 \Psi_{2}^{\circ}\Edt'\Omega^{\circ}+\rho^2\bar{\rho}\Phi_{21}^{\circ}+2
 \rho^3\bar{\rho}\Edt'\Phi_{11}^{\circ}+3\rho^4\bar{\rho}\Phi_{11}^{\circ}
 \Edt' \Omega^{\circ} \nonumber \\ \Psi_{4} & = & \rho\Psi_{4}^{\circ}+\rho^2
 \Edt'\Psi_{3}^{\circ}+\frac{1}{2}\rho^3\bigl(\Edt'^2\Psi_{2}^{\circ}+2
 \Psi_{3}^{\circ}\Edt'\Omega^{\circ}\bigr)+\frac{1}{2}\rho^4\bigl(
 \Psi_{2}^{\circ}\Edt'^2\Omega^{\circ}+3\Edt'\Omega^{\circ}\Edt'
 \Psi_{2}^{\circ}\bigr) \nonumber \\ & & +\frac{3}{2}\rho^5\Psi_{2}^{\circ}
 (\Edt'\Omega^{\circ})^2+\rho^2\bar{\rho}\Edt'\Phi_{21}^{\circ}+\rho^3
 \bar{\rho}\bigl(\Edt'^2\Phi_{11}^{\circ}+\Phi_{21}^{\circ}\Edt'
 \Omega^{\circ}\bigr) \nonumber \\ & & +\rho^4\bar{\rho}\bigl(
 \Phi_{11}^{\circ}\Edt'^2\Omega^{\circ}+3\Edt'\Omega^{\circ}\Edt'
 \Phi_{11}^{\circ}\bigr)+3\rho^5\bar{\rho}\Phi_{11}^{\circ}(\Edt'
 \Omega^{\circ})^2 \nonumber \\ \Phi_{11} & = & \rho^2\bar{\rho}^2
 \Phi_{11}^{\circ} \nonumber \\ \Phi_{21} & = & \rho\bar{\rho}^2
 \Phi_{21}^{\circ}+\rho^2\bar{\rho}^2 \Edt' \Phi_{11}^{\circ}+\rho^3
 \bar{\rho}^2\Phi_{11}^{\circ}\Edt'\Omega^{\circ} \nonumber \\ 
 \Phi_{22} & = & \rho\bar{\rho}\Phi_{22}^{\circ}+\rho^2\bar{\rho}\bigl(\Edt'
 \overline{\Phi}_{21}^{\circ}-\frac{1}{2}\Pht'\Phi_{11}^{\circ}\bigr)
 +\rho\bar{\rho}^2\bigl(\Edt\Phi_{21}^{\circ}-\frac{1}{2}\Pht'\Phi_{11}^{\circ}
 \bigr) \nonumber \\Ê& & +\rho^3\bar{\rho}\overline{\Phi}_{21}^{\circ}\Edt'
 \Omega^{\circ}+\frac{1}{2}\bigl(\Edt\Edt'\Phi_{11}^{\circ}+\Edt'\Edt
 \Phi_{11}^{\circ}\bigr)-\rho\bar{\rho}^3\Phi_{21}^{\circ}\Edt\Omega^{\circ}
 \nonumber \\ & & +\rho^3\bar{\rho}^2 \Edt'\Omega^{\circ}\Edt\Phi_{11}^{\circ}
 -\rho^2\bar{\rho}^3\Edt\Omega^{\circ}\Edt'\Phi_{11}^{\circ}-\rho^3
 \bar{\rho}^3\Phi_{11}^{\circ}\Edt\Omega^{\circ}\Edt'\Omega^{\circ}.
\end{eqnarray}
The remaining Ricci and Bianchi equations are
\begin{eqnarray}
 \Pht'\rho'^{\circ}-\Edt\kappa'^{\circ} & = & (2\Omega^{\circ}\rho'^{\circ}
 +\Psi_{2}^{\circ}-\overline{\Psi}_{2}^{\circ})\Lambda \nonumber \\
 \Edt'\kappa'^{\circ} & = & -\Psi_{4}^{\circ} \nonumber \\ \Edt'
 \rho'^{\circ} & = & -\Omega^{\circ}\kappa'^{\circ}-\Psi_{3}^{\circ}
 \nonumber \\ \Edt\Psi_{2}^{\circ} & = & 2\overline{\Phi}_{21}^{\circ}
 \nonumber \\ \Edt\Psi_{3}^{\circ}+\Edt'\overline{\Psi}_{3}^{\circ}-\Pht'
 (\Psi_{2}^{\circ}+\overline{\Psi}_{2}^{\circ}) & = & 2\Phi_{22}^{\circ}.
 \label{RicBian2}
\end{eqnarray}
The commutators become
\begin{eqnarray}
 \Bigl[\Ph,\Edt\Bigr] & = & 0 \nonumber \\ \Bigl[\Ph,\Edt'\Bigr] & = & 0
 \nonumber \\Ê\Bigl[\Ph,\Pht'\Bigr] & = & -\Bigl(\frac{1}{2}\rho^2
 \Psi_{2}^{\circ}+\frac{1}{2}\bar{\rho}^2\overline{\Psi}_{2}^{\circ}+
 \rho^2\bar{\rho}\Phi_{11}^{\circ}+\rho\bar{\rho}^2\Phi_{11}^{\circ}+
 \Lambda\bigl(\frac{1}{\rho}+\frac{1}{\bar{\rho}}\bigr)\Bigr)\Ph
 \nonumber \\ \Bigl[\Pht',\Edt\Bigr] & = & \Bigl(-\frac{\bar{\kappa}'^{\circ}}
 {\bar{\rho}}+\overline{\Psi}_{3}^{\circ}+\frac{1}{2}\bar{\rho}\Edt
 \overline{\Psi}_{2}^{\circ}-\frac{1}{2}\bar{\rho}^2\overline{\Psi}_{2}^{\circ}
 \Edt\Omega^{\circ}+\rho\overline{\Phi}_{21}^{\circ}+\rho\bar{\rho}\Edt
 \Phi_{11}^{\circ} \nonumber \\Ê& & -\rho\bar{\rho}^2\Phi_{11}^{\circ}\Edt
 \Omega^{\circ}\Bigr)\Ph+q\bigl(\bar{\kappa}'^{\circ}-\Lambda\Edt\Omega^{\circ}
 \bigr)\nonumber \\ \Bigl[\Pht',\Edt'\Bigr] & = & \Bigl(-\frac{\kappa'^{\circ}}
 {\rho}+\Psi_{3}^{\circ}+\frac{1}{2}\rho\Edt'\Psi_{2}^{\circ}+\frac{1}{2}\rho^2
 \Psi_{2}^{\circ}\Edt'\Omega^{\circ}+\bar{\rho}\Phi_{21}^{\circ}+\rho
 \bar{\rho}\Edt'\Phi_{11}^{\circ} \nonumber \\ & & +\rho^2\bar{\rho}
 \Phi_{11}^{\circ}\Edt'\Omega^{\circ}\Bigr)\Ph+p\bigl(\kappa'^{\circ}+
 \Lambda\Edt'\Omega^{\circ}\bigr) \nonumber \\ \Bigl[\Edt,\Edt'\Bigr] & = &
 \Bigl(\frac{\bar{\rho}'^{\circ}}{\bar{\rho}}-\frac{\rho'^{\circ}}{\rho}+
 \frac{\rho}{2}\bigl(\frac{1}{\rho}+\frac{1}{\bar{\rho}}\bigr)\Psi_{2}^{\circ}
 -\frac{\bar{\rho}}{2}\bigl(\frac{1}{\rho}+\frac{1}{\bar{\rho}}\bigr)
 \overline{\Psi}_{2}^{\circ}+\Omega^{\circ}(\rho\bar{\rho}\Phi_{11}^{\circ}
 -\Lambda)\Bigr)\Ph \nonumber \\Ê& & +\Omega^{\circ}\Pht'+p\bigl(
 \rho'^{\circ}+\Omega^{\circ 2}\Lambda\bigr)-q(\bar{\rho}'^{\circ}
 +\Omega^{\circ 2}\Lambda\bigr) \label{Commutators}
\end{eqnarray}

\subsection{Conclusions}

The most obvious application of these results is when searching for
exact non-vacuum solutions of Einstein's field equations using the GHP 
integration procedure suggested by Held \cite{Held1}, \cite{Held2} and
developed by Edgar and Ludwig \cite{Edgar}, \cite{EL1}, \cite{EL2}. It is
worth noting that among the remaining Ricci- and Bianchi equations
(\ref{RicBian2}) we can view the second one as the definition of
$\Psi_{4}^{\circ}$, the third as the definition of $\Psi_{3}^{\circ}$,
the fourth as the definition of $\Phi_{21}^{\circ}$ and the fifth as the
definition of $\Phi_{22}^{\circ}$. This leaves only a system of equations
consisting of the equations (\ref{rOmegaeqns}), (\ref{Commutators}) and
the first equation of (\ref{RicBian2}) for the unknown functions
$\frac{1}{\rho}+\frac{1}{\bar{\rho}}$, $\Omega^{\circ}$, $\kappa'^{\circ}$,
$\rho'^{\circ}$, $\Psi_{2}^{\circ}$, $\Phi_{11}^{\circ}$ and the constant
$\Lambda$. Of these, $\frac{1}{\rho}+\frac{1}{\bar{\rho}}$,
$\Phi_{11}^{\circ}$ and $\Lambda$ are real while $\Omega^{\circ}$ is
purely imaginary and the others are complex.

There are however two important points to keep in mind.
\begin{itemize}
 \item The GHP integration program requires that the commutators are 
       applied to four real $(0,0)$-weighted quantities and one 
       non-trivially weighted complex quantity. Only then is all the 
       information extracted from the commutators.
       
 \item We have not yet imposed Einstein's field equations. When we do
       this some of the equations taken as definitions of the Ricci 
       components will actually become constraints e.g., if we impose 
       the condition that the spacetime be vacuum we obtain the extra
       conditions 
       \begin{eqnarray*}
        \Edt\Psi_{2}^{\circ} & = & 0 \\ \Edt\Psi_{3}^{\circ}-
        \Pht'\Psi_{2}^{\circ} & = & 0.
       \end{eqnarray*}
       Also, impositions of other features such as symmetries, special 
       Petrov types etc., may result in extra constraint equations.
\end{itemize}

Another important application is to quasi-local momentum in these 
spacetimes. Recall that a symmetric spinor $L_{ABCA'}$ is said to be
a Lanczos potential of $\Psi_{ABCD}$ if it satisfies the Weyl-Lanczos 
equation i.e.,
$$
 \Psi_{ABCD}=2\nabla_{(A}{}^{A'}L_{BCD)A'}
$$
This equation can be translated into GHP-formalism (see e.g.,
\cite{Andersson}) and if we look for solutions satisfying
$$
 L_{ABCA'}o^{A'}=0
$$
the GHP Weyl-Lanczos equations are easy to $\rho$-integrate. Such
Lanczos potentials turn out to be very useful when looking for metric 
asymmetric curvature-free connections in these spacetimes. Such a 
connection have been used in the Kerr spacetime to construct 
quasi-local momentum. For details of this construction and some 
generalizations, see \cite{BL}, \cite{Harnett}, \cite{Bergqvist}, 
\cite{Andersson} and \cite{AE1}. This application is further 
developed in \cite{Andersson2}.

\section*{Acknowledgements}

Special thanks are due to docent S. Brian Edgar for helpful 
suggestions and discussions.

\end{document}